# PZT Optical Memristors

Chenlei Li[1+], Hongyan Yu[2+], Tao Shu[1+], Yueyang Zhang[1], Chengfeng Wen[1], Hengzhen Cao[1], Jin Xie[1], Hanwen Li[1], Zixu Xu[1], Gong Zhang[1], Zejie Yu[1], Huan Li[1], Liu Liu[1], Yaocheng Shi[1], Feng Qiu[2*], Daoxin Dai[1,3*]

[1]State Key Laboratory for Extreme Photonics and Instrumentation, College of Optical Science and Engineering, International Research Center for Advanced Photonics, Zhejiang University, Zijingang Campus, Hangzhou 310058, China

[2] Hangzhou Institute for Advanced Study, University of Chinese Academy of Sciences, Hangzhou 310024, Zhejiang Province, China

[3] Intelligent Optics and Photonics Research Center, Jiaxing Research Institute, Zhejiang University, Jiaxing 314000, China

[+]These authors contributed equally to this work.
*Corresponding Author: dxdai@zju.edu.cn; A-photonics@outlook.com

## Abstract

Optical memristors represent a monumental leap in the fusion of photonics and electronics, heralding a new era of applications from neuromorphic computing to artificial intelligence. However, current technologies are hindered by complex fabrication, limited endurance, high optical loss or low modulation depth. For the first time, we reveal optical non-volatility in thin-film Lead Zirconate Titanate (PZT) by electrically manipulating the ferroelectric domains to control the refractive index, providing a brand-new routine for optical memristors. The developed PZT optical memristors offer unprecedented advantages more than exceptional performance metrics like low loss of <2 dB/cm, high precision exceeding 6-bits, large modulation depth with an index change as large as $4.6\times10^{-3}$. Additionally, these devices offer impressive stability, maintaining minimal wavelength variation for over three weeks and enduring more than 10,000 cycles, and require a mere 0.8 pJ of energy for non-volatile operation. The wafer-scale sol-gel fabrication process also ensures compatible with standardized mass fabrication processes and high scalability for photonic integration. Specially, these devices also demonstrate unique functional duality: setting above a threshold voltage enables non-volatile behaviors, below this threshold allows volatile high-speed optical modulation. This marks the first-ever optical memristor capable of performing high-speed (48 Gbps) and energy-efficient (450 fJ/bit) signal processing and non-volatile retention on a single platform, and is also the inaugural demonstration of scalable functional systems. The PZT optical memristors developed here facilitate the realization of novel paradigms for high-speed and energy-efficient optical interconnects, programmable PICs, quantum computing, neural networks, in-memory computing and brain-like architecture.

## Introduction

Memristors, with the ability to modulate signals actively and hold memory states akin to synaptic activity in the mammalian brain, have opened new avenues to non-volatile memory and neuromorphic computing, significantly improving energy efficiency and integration density while bypassing the von Neumann bottleneck [1]. Inspired by the remarkable capabilities and technological promise of electronic memristors, the photonics community has been striving to develop an optical counterpart called optical memristors, which are capable of modulating the light amplitude/phase and holding the state non-volatilely[2]. Optical memristors represent a monumental leap in the fusion of photonics and electronics, heralding a new era of high-speed energy-efficient information processing that promises to transcend the constraints of traditional architectures.

Significant efforts haves been made towards developing optical memristors to create new paradigms in photonic integrated circuits (PICs), and catalyze the emergence of new applications, including high-efficiency in-memory computing[3-6], brain-inspired architectures[7], post-fabrication trimming of PICs for calibrating the operation of sophisticated programmable optical processors with fabrication-sensitive structures. Among them, phase-change materials (PCMs) [8-16] have received substantial attention owing to some favorable property, while the difficulty in the inherent control mechanism for the amorphous/crystalline phase change might hinder the wide application in large-scale integration[2]. Microelectromechanical systems (MEMS)[17,18], magneto-optics[19] and charge-based[20,21] optical memristors are interesting but still suffer from the fabrication complexity or inability for multi-level tuning, and few experimental demonstrations have been reported so far[2]. Ferroelectric material is another option allows storage of optical information by switching non-volatile ferroelectric domains through applying an external electric field[22,23]. In ref. [23], the Pockels coefficient was controllable, and hence the refractive index changed under a direct current (DC) bias applied to the thin-film $BaTiO_3$ (BTO). However, the heterogeneous integration of thin-film BTO required a few complicated processes including molecular beam epitaxy, direct wafer bonding and wafer grinding, while the modulation depth of $2\times10^{-3}$ and the propagation loss of 4.8 dB/cm for the BTO waveguides leave space for further refinement. Furthermore, an external DC bias was needed continuously and thus maintaining the state still requires power supply, resulting in continuous power consumption at the idle state. Therefore, more efforts are expected to explore a brand-new routine for optical memristors in terms of optical performance, switching speed, energy, fabrication process, as well as the capability for scaling to system-level integration[2].

To address these challenges, we propose optical memristors based on thin-film PZT for the first time, which precisely control the refractive index directly by electrically manipulating the ferroelectric domain polarization, no bias voltage required. Specifically, our PZT optical memristors have a unique functional duality related with a threshold voltage $V_{th}$ (which has not yet been available in reported platforms): operating above $V_{th}$ enables non-volatility with permanent changes in domain polarization, while working below $V_{th}$ achieves fast volatile optical switching/modulation with high-efficiency Pockels effect. First, for the non-volatile operation, the developed PZT optical memristors offer unprecedented advantages more than exceptional performance metrics low loss of <2 dB/cm, high precision exceeding 6-bits, large modulation depth with an index change as large as $4.6\times10^{-3}$, high stability over the period of more than three weeks, great endurance of over 10000 cycles at the frequency of 1kHz with pJ-level non-volatile switching energy and zero maintaining power. Second, for the volatile operation, we efficiently achieve high speed of 48 Gbps and high energy efficiency of 450 fJ/bit for signal processing with high modulation efficiency of 42 pm/V. Finally, the wafer-scale sol-gel fabrication process ensures CMOS compatibility and scalability, making it ideal for on-chip photonic integration in wafer-scale. This advancement enable the successful development of crucial building blocks such as microring resonators (MRRs), Mach-Zehnder interferometers (MZIs), and Fabry-Pérot (FP) cavities for optical memristors, showing great potential as the essential elements with versatile functionality for advancing large-scale, on-chip programmable photonic systems[24]. Furthermore, we demonstrate the scalability of PZT optical memristors for developing PICs, featuring a memristive optical switch matrix and an adaptive memristive optical cavities for spectral manipulation. This work marks the first-ever optical memristor capable of performing high-speed signal processing and non-volatile retention on a single platform, and presents the inaugural demonstration of scalable functional systems with electrical-controlled optical memristors. Our finding underscores the great promise for wide applications in high-speed and energy-efficient optical interconnects[25], programmable PICs[26], quantum computing[27,28], neural networks[4,29], in-memory computing[6] and brain-like computing[30].

# Results

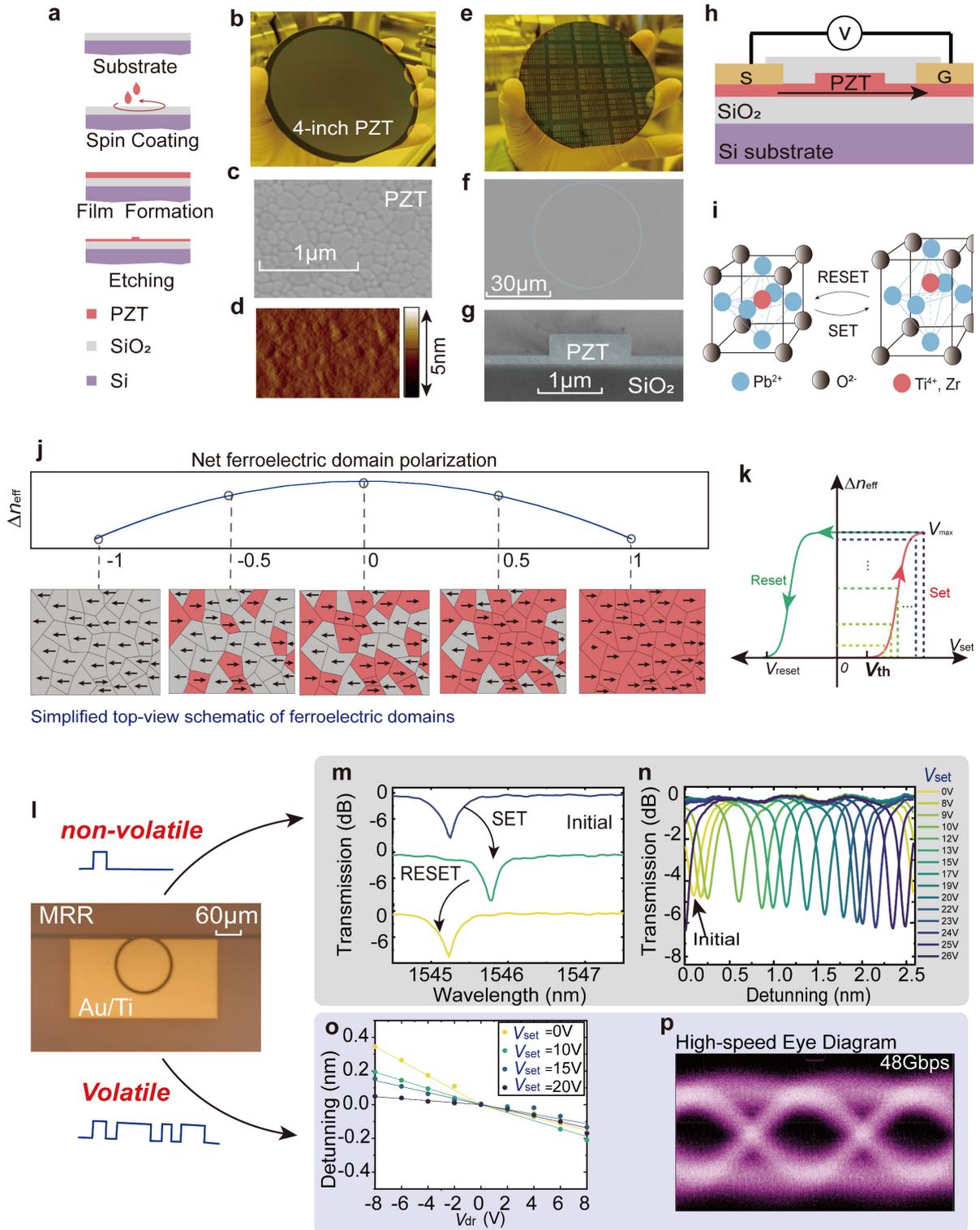

**Fig. 1 | Optical PZT memristor. a.** Wafer-scale sol-gel fabrication processes of PZT Optical Memristors. **b.** Photograph of the spin-coating wafer demonstrating uniform and defect-free bonding. **c.** Scanning electron microscopy (SEM) picture of top view of the thin-film PZT (thickness~ 300nm). **d.** AFM image of the thin-film PZT. **e.** Photograph of a 4-inch wafer including the fabricated PZT optical memristors patterned by using an ultraviolet stepper lithography system. **f.** Colorized SEM of PZT MRR (blue). **g.** SEM picture of the cross section of a PZT ridge waveguide. **h.** Schematic diagram of the ridge waveguide cross-section and the poling process. **i.** PZT unit cell representation before and after poling. **j.** The dependence of the refractive index changes $\Delta n_{\text{eff}}(P_v)$ against the net ferroelectric domain polarization, and the corresponding simplified top-view schematic of ferroelectric domains in PZT. **k.** Hysteresis loop of $V_{\text{set}}/V_{\text{reset}}$

against $\Delta n_{\text{eff}}$. **l** Optical microscopy image of an MRR with a bending radius of 60 μm. **m.** Single-cycle demonstration of one distinct non-volatile state, inset: procedure of setting protocol. **n.** Measured transmissions at the through port of the MRR-based PZT optical memristor with different setting voltages $V_{\text{set}}$. **o.** Measured resonance-wavelength detuning of the MRR-based PZT optical memristor set with different setting voltages $V_{\text{set}}$ as the driving voltage $V_{\text{dr}}$ varies. **p.** Eye diagrams for NRZ modulation at the data rate of 48 Gbps with a driving voltage $V_{\text{dr}}$ of 3 V.

The fabrication of thin-film PZT wafers and optical waveguides was carried out with the sol-gel process, as illustrated schematically in Fig. 1a-g (see more details in Methods). The 4-inch PZT wafer was realized with a top-surface roughness of 0.7 nm and a thickness non-uniformity less than 5 nm, as shown in Fig. 1b-d. All the following processes, including the lithography, dry etching, and by-product cleaning processes, were optimized for achieving PZT optical waveguides with smooth sidewalls/top-surfaces, enabling high-quality building blocks and multi-functional photonic integrated circuits (PICs), as depicted in (Fig. 1e-g). Such a wafer-scale sol-gel fabrication process developed for PZT PICs ensures excellent CMOS compatibility and high scalability as well as low-cost production. The cross-section of the fabricated PZT waveguide is shown in Fig. 1e, where the poling electrodes are strategically placed on the slab and on both sides of the ridge.

As a ferroelectric material, PZT possesses a perovskite crystal structure. Initially, the crystal is in a paraelectric phase, with positive and negative charge centers aligned at the body center (see the left part of Fig. 1i). When a poling voltage $V_{\text{pol}}$ is applied to the thin-film PZT, Zr and $Ti^{4+}$ ions shift towards one of the eight vertices, causing a displacement of the positive and negative charge centers. This shift introduces the domain reorientation of PZT, enhancing its polarizability. As a result, these compositions exhibit excellent dielectric and ferroelectric properties[31] (Fig. 1i). After removing the external electric field, the ferroelectric domain inside the thin-film PZT maintains a certain degree of polarization, known as *remanent polarization*[32]. The refractive index $n_{\text{eff}}(P_v)$ of thin-film PZT is related to the remnant polarization in PZT[33], i.e.,

$$n_{\text{eff}}(P_v) = n_{\text{eff0}} + n_1 \frac{P_v^2}{P_0^2}, \tag{1}$$

where $n_{\text{eff0}}$ is the refraction index of thin-film PZT with random polarization, $n_1$ is the relative refractive index, $P_v$ is the net ferroelectric domain polarization, and $P_0$ is the ferroelectric domain polarization on the boundary of the thin film. According to equation (1), a different ratio of the mixed-polarization domains in PZT, depending on the setting voltage $V_{\text{set}}$, would result in a different refractive index $n_{\text{eff}}(P_v)$, as described in Fig. 1j (Supplementary section **1**). Therefore, non-volatile "SET" process can be controlled by applying different setting voltage $V_{\text{set}}$, and thus achieve different $n_{\text{eff}}(P_v)$. Similarly, applying a reverse voltage can reset the net ferroelectric domain polarization in PZT to its *initial* state. The relationship between $\Delta n_{\text{eff}}$ against $V_{\text{set}}$ is depicted by the hysteresis loop of PZT given in Fig. 1k.

Notably, the device is revealed to exhibit a unique functional duality related with a threshold voltage $V_{\text{th}}$. To achieve the non-volatile effect, the applied voltage should be higher than $V_{\text{th}}$ to induce notable changes in ferroelectric domain polarization. On the other hand, when the voltage is less than $V_{\text{th}}$, the weak electric field does not induce notable changes in ferroelectric domain polarization, and thus the refractive index is modulated electrically via the Pockels effect in PZT volatilely, providing an effective routine for realizing high-speed electro-optic (EO) modulation or switching without triggering the non-volatile effect.

To show the unique functional duality, we employed a thin-film PZT MRR for realizing multi-level non-volatile operation and high-speed volatile operation (Fig. 1l). With an electrode spacing of 4 μm, the threshold voltage for this MRR is determined experimentally as ~8 V. In this experiment, we control the ferroelectric domain by applying 10 voltage–pulses with the frequency of 1 kHz to the devices, where $V_{\text{set}}$ is varied from 8 V to 26 V. The total non-volatile switching time is ~10 milliseconds, and accordingly the non-volatile switching energy is calculated as 0.8-2.6 pJ (Supplementary section **2**). Fig. 1m shows a single-cycle demonstration of one distinct non-volatile state and varying the setting voltage $V_{\text{set}}$ results in multi-level non-volatile resonance-wavelength detuning, as illustrated in Fig. 1n. Piezoresponse Force Microscopy (PFM) images intuitively show the change of the domain polarizations in the thin-film PZT waveguide with different setting voltages $V_{\text{set}}$ in Supplementary section **3**. For high-speed volatile operation, the EO modulation efficiency of the PZT waveguides and its dependence on $V_{\text{set}}$ were firstly characterized by measuring the resonance-wavelength detuning of the fabricated MRR against the driving voltage $V_{dr}$ when it was set with $V_{set}$= 0, 10, 15, and 20 V. Here $V_{dr}$ was varied from −8 V to 8 V, staying below the threshold, as shown in Fig. 1o. When $V_{dr}$ was negative (i.e., opposite to the poling direction), the measured modulation efficiency was about 42, 25, 12, and 6 pm/V with $V_{set}$= 0, 10, 15, and 20 V, respectively. When $V_{dr}$ was positive, the modulation efficiency became 26, 33, 20, and 23 pm/V, respectively. Such asymmetry of the $V_{set}$-dependence for the modulation efficiency is attributed to the special mechanism in PZT, as expected by our theoretical prediction in Supplementary section **1**. Then high-speed EO modulation for NRZ signals was demonstrated by using the setup described in method. As given in Fig. 1p, the recorded NRZ eye-diagram with a driving voltage $V_{dr}$ of 3 V at 48 Gbps, the capacitance is approximately 200 fF and the corresponding electrical energy consumption is estimated to be ~450 fJ/bit for the 48 Gbps data transmitting.

Further measurements confirmed the fabricated PZT waveguide had a low propagation loss of ~1.9 dB/cm and low temperature sensitivity of ~30 pm/°C (Supplementary sections **4** and **5**). Additionally, we also investigated the DC drift of the thin-film PZT waveguides (Supplementary Section **6**), finding high stability with a negligible phase drift of less than $0.05\pi$. The ability to address the issue of DC drifting, which often challenges other EO modulators or switches fabricated on e.g. thin-film lithium niobate[34], is very promising for various applications with optical switching and modulation. Consequently, thin-film PZT provides a promising option as a brand-new platform for developing on-chip optical memristors, and thus we present several representative building blocks for various applications in the following sections.

## Building blocks

### MZI-based PZT optical memristors

MZIs are crucial fundamental components for photonic systems on a chip, encompassing the versatile functionality of e.g. optical modulation, optical switching, variable optical attenuation, optical filtering, variable optical splitting for optical interconnects and optical computing [35,36]. For example, large-scale photonic switches are typically assembled from 2×2 MZI switches interconnected into networks or arrays of various topologies, including cross-bar, Benes, and path-independent loss (PI-Loss), along with their numerous variants [29,35,37]. However, developing large-scale PICs presents significant challenges, particularly in calibrating a large number of MZI elements due to random phase errors in their initial states, which often require complex processes and additional power consumption. In this section, we demonstrate MZI-based PZT optical memristors for optical switching, Fig. 2a-b, highlighting their potential to simplify the calibration and non-volatile reconfigurability as well as volatile high-speed optical signal processing.

The present MZI-based PZT optical memristor, consists of two imbalanced arms and two 2×2 multimode interferometers with a splitting ratio of 50%: 50%. Electrodes used for applying a lateral electric field were placed on both sides of these two arm waveguides, forming the phase shifters with the length of e.g. ~1.0 mm. Initially, the poling process is implemented by applying ten short electrical pulses at the frequency of 1 kHz in the direction shown in Fig. 2a. Fig. 2c shows the measured non-volatile quasi-continuous tuning of the notch wavelength as $V_{set}$ increases with 20 levels from 10 V to 32 V, while accordingly the energy required ranges from 1 pJ to 3.6 pJ.

To further demonstrate the functional duality of thin-film PZT, the MZI operated in a push-pull configuration shown in Fig. 2a for high-speed optical signal processing with a driving voltage $V_{dr}$ lower than $V_{th}$. Fig. 2d shows the measurement results when 4 $V_{pp}$ (Voltage peak-peak) was applied with 10 MHz frequency (Methods), indicating that the device is switchable in nano-second scale. Fig. 2e presents the measured transmissions at the cross port of the MZI with 1-mm-long phase shifters when operating with $V_{dr}$= 0 and 5.0 V, respectively, showing that a phase-shifting of $\pi$ was achieved, and correspondingly the modulation efficiency $V_\pi L$ is about 0.5 V·cm. The estimated Pockels coefficient of the post-poling PZT is as high as 120 pm/V, which is over four times higher than that of lithium niobate[38], thus paving a revolutionary routine for realizing high-efficiency and high-speed EO switching/modulation.

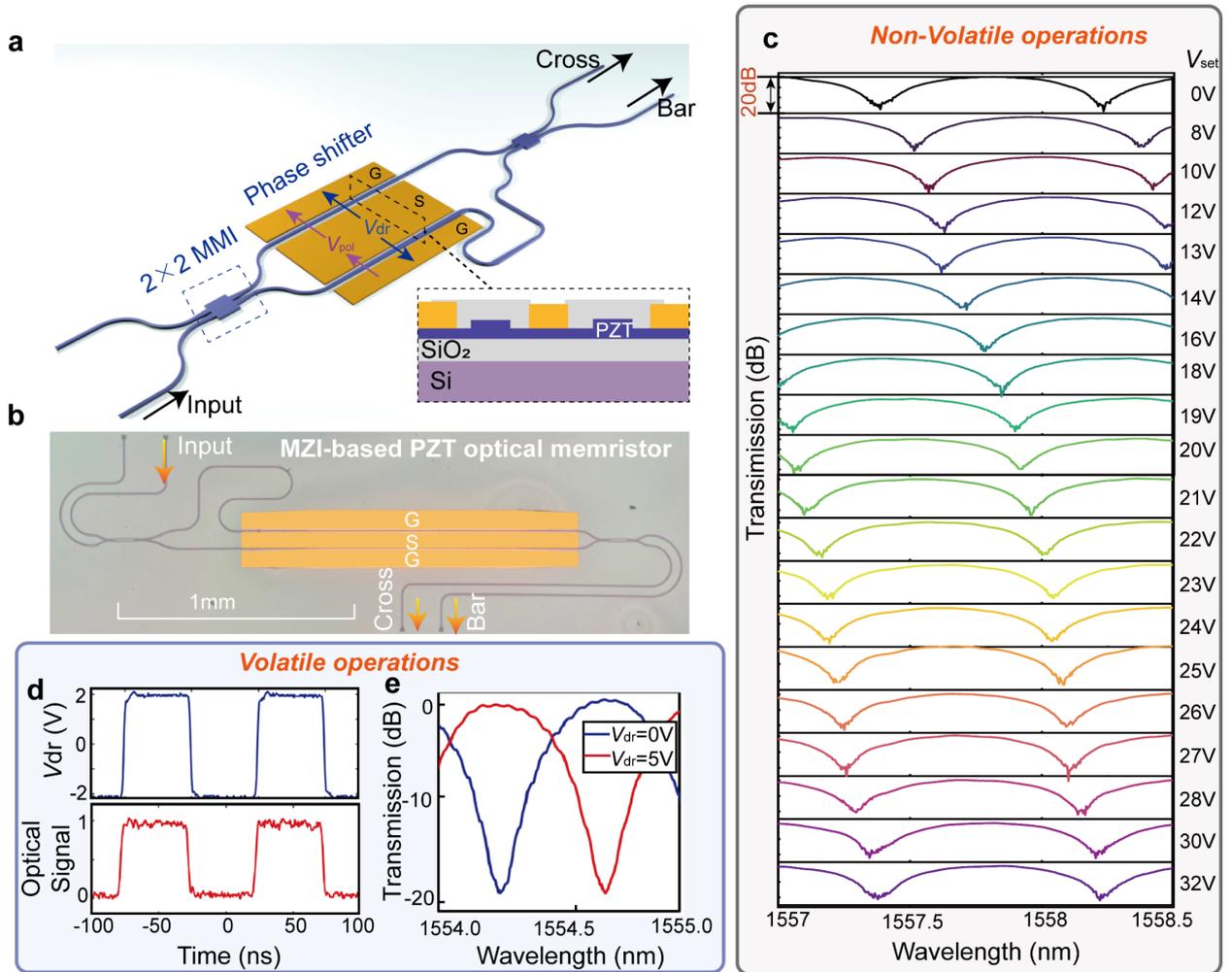

**Fig. 2 | The present MZI-based PZT optical memristors. a.** Schematic of the MZI-based 2×2 PZT optical memristors, consisting of two imbalanced arms with 1-mm-long phase shifters and two 2×2 multimode interferometers with a splitting ratio of 50%: 50%; the purple arrow represents the direction of the poling electric field; the blue arrow represents the direction of the applied drive voltage $V_{dr}$. **b.** Optical microscopy image of the fabricated 2×2 MZI optical memristors. **c.** Quasi-continuous tuning of the wavelength notch with the setting voltage $V_{set}$ varying from 8 V to 32 V. Before the setting voltage is applied to set any state, this optical memristor was re-initialized by applying a high poling voltage $V_{pol}$ of 80 V. **d.** Real-time optical modulation measurements. Normalized output intensity from the through port when the device operates with a drive voltage $V_{dr}$ of 4 $V_{pp}$ at the frequency of 10 MHz. The upper one represents the input electrical signals, and the bottom one represents the switched optical signals. **e.** Measured transmissions at the cross port when operating with $V_{dr}$ =0 and 5.0 V, respectively.

## FP-cavity-based PZT optical memristors

Optical cavities have been recognized as another key element for realizing various functionality (including optical filtering, modulation or switching) due to the great advantages in structure flexibility, footprint compactness and energy efficiency [39-41]. Unfortunately, the resonance wavelengths are often extremely susceptible to the fabrication variations and thus one often needs to thermally or electrically tune the resonance wavelength for alignment with additional high-power consumption[42].

In this section, we propose a 1×2-FP-cavity-based PZT optical memristor, as shown in Fig. 3a-b (more design details in Supplementary section 7). Note that the present device has a short cavity length, enabling a free spectral range (FSR) as large as 8 nm, which is much larger than regular MRRs on the same chip, helping to clearly observe the multi-level non-volatility revealed here. With this PZT optical memristor, we achieved quasi-continuous 64-level (6-bit) non-volatile setting by varying the voltage $V_{set}$ from 10 V to 36 V, as shown in Fig. 3c-d. The maximum wavelength-detuning reaches up to 3.0 nm ,

which is greater than those reported optical memristors and the corresponding non-volatile refractive-index change was approximately $4.6 \times 10^{-3}$ (Supplementary section **8**).

Such quasi-continuous non-volatility enables the phase-shift manipulating of photonic devices precisely. Meanwhile, the Q-factor of the resonance peaks remains almost unchanged, indicating that the multi-level non-volatility enabled by manipulating the domain polarization in thin-film PZT does not introduce a notable influence on the waveguide's propagation loss. More levels of non-volatility could be achieved by further refining the setting voltage $V_{set}$. Fig. 3e demonstrates the remarkable repeatability and stability of the FP-cavity-based optical memristor by characterizing five target states set with $V_{set}$=30, 21, 23, 25, and 36 V as examples, respectively. For each target state, the following cycle process was repeated 20 times (Supplementary section **9**): (i) setting the target state; (ii) measuring the resonance wavelength; (ii) erasing to the initial state. These twenty resonance wavelengths corresponding to the same setting voltage $V_{set}$ have very excellent consistency and repeatability with a minor wavelength variation of less than ±0.05 nm, validating the great potential of PZT optical memristors for creating highly reliable and multi-level non-volatile photonic elements, more details are shown in Supplementary section **10**. To showcase the high endurance of our PZT optical memristors, we utilized an arbitrary function generator to cycle between SET and RESET pulses at the frequency of 2 Hz, with the pulse-series set at 30 V for SET and 150 V for RESET, while each pulse has a duration of 1.0 ms. Optical transmission data was manually recorded every two minutes throughout the 90-minute experiment, as depicted in Fig. 3f. Remarkably, even after more than 10,000 cycles, the device exhibited no signs of degradation.

The functional duality of the FP-cavity-based PZT memristor was also showcased in experiments. When operating below $V_{th}$, this PZT memristor demonstrated its availability for high-speed optical switching/modulation. In particular, temporal responses of the device were measured with different modulation frequencies of 1 Hz, 100 kHz, and 5 MHz, while the driving voltage was chosen as 4 $V_{pp}$, as shown in Fig. 3g-j. The optical signal was effectively modulated across different frequencies, showcasing the device's versatility for high-speed switching with minimal electro-optic relaxation effects, which is a crucial issue for low-frequency or long-timescale applications, like ubiquitous DC bias in modulators[43]. As shown in Fig. 3i, the switching exhibited a fast rising/falling time of around 2.5 ns, primarily limited by the measurement setup in the lab. Intrinsically, the switching speed can be achieved in sub-nanosecond scale. Fig. 3k demonstrates the measured static spectral responses at the through port of this FP cavity when operated with different driving voltages $V_{dr}$ of −10, 0, and +10 V, respectively. The EO modulation efficiencies for negative and positive biases were different with the measured values of 30 pm/V and 20 pm/V, respectively, showing the asymmetry similar to the MRR-based PZT optical memristors mentioned above. By judiciously choosing the appropriate ferroelectric domain polarization and bias voltage, the effective Pockels coefficient can be enhanced, which is advantageous for developing high-performance EO modulators as desired for high-speed optical processing systems.

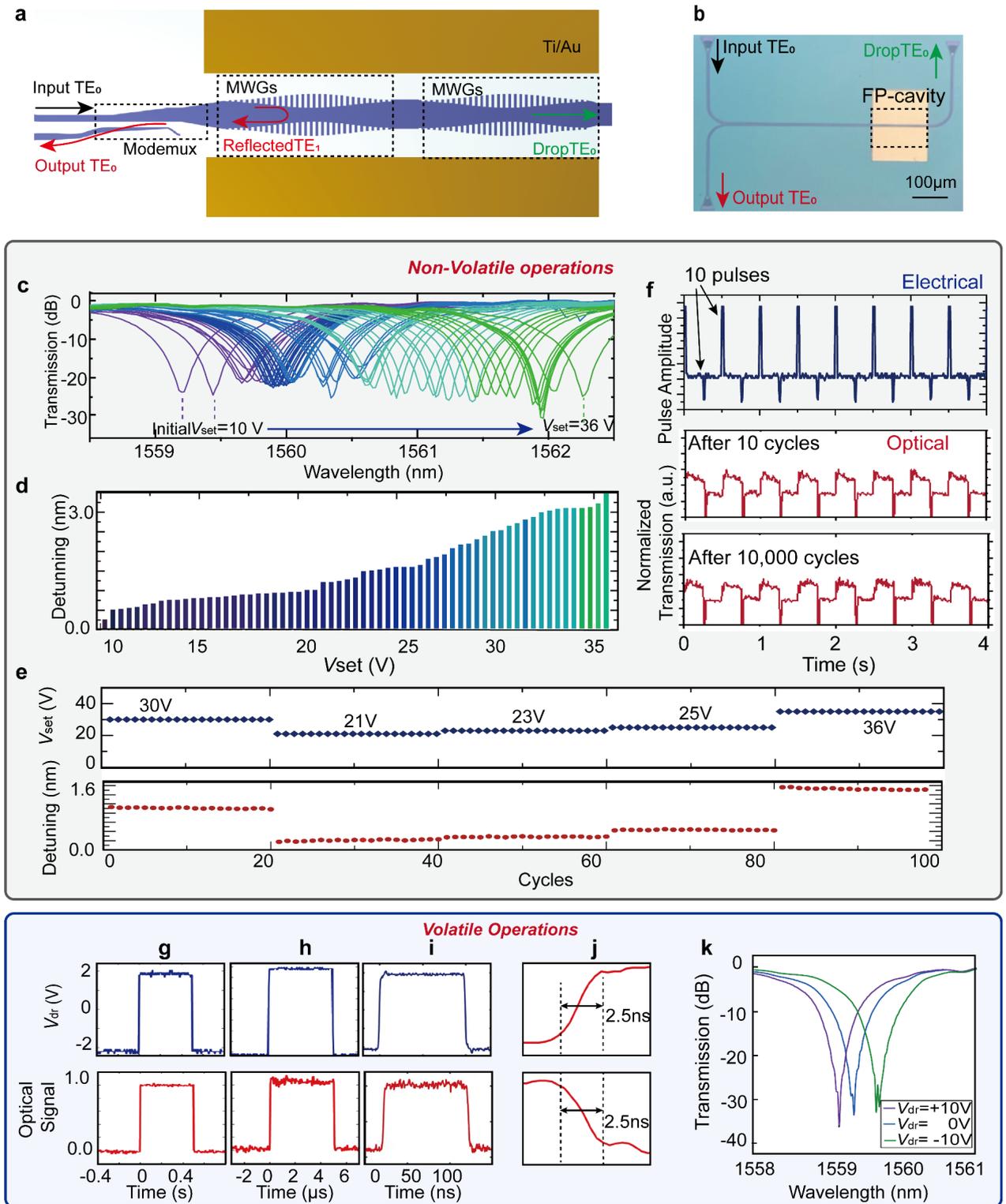

**Fig. 3 | The present FP-cavity-based PZT optical memristor. a.** Schematic configuration. MWGs, multimode waveguide gratings, Modemux, mode multiplexers. **b.** Optical microscopy image of the fabricated device. **c.** Measured transmission spctral responses at the through port with different setting voltages $V_{set}$. **d.** 64-level setting process (detuning with 1559.2 nm) induced by applying different setting voltages $V_{set}$. **e.** Five distinct non-volatile states, corresponding to the setting voltages $V_{set}$ of 30, 21, 23, 25, and 36 V. For each non-volatile state, there are 20 cycles measured for the resonance-wavelength detuning. **f,** Time domain measurements after 10 and 10000 cycles, showing that the optical memristor operates well without any degradation. s. Real-time optical modulation measurements: **g-j**: Normalized output optical intensity at the through port when operating with rectangular driving-voltage signals ($V_{dr}=4V_{pp}$) at the frequency of 1 Hz (**g**), 100 kHz (**h**) and 5 MHz (**i**). **j.** The rising/falling edge of the switched optical signals. **k.** Measured spectral responses at the through port of the FP cavity when operating with different driving voltages $V_{dr}$ of −10, 0, and +10 V, respectively.

# Multifunctional Photonic Circuits Enabled by Scalable PZT Optical Memristors

Previously reported optical memristors were still limited to the implementations of individual elements (such as MZIs and MRRs) [12,23], and have not demonstrated their potential scalability for multifunctional photonic systems on a chip with large-scale integration, which is essential for practical applications. Here, we make the efforts to demonstrate the scalability and versatility of our PZT optical memristors with the following two representative architectures, including a memristive optical switch matrix and an adaptive memristive multi-channel optical cavities for spectral manipulation.

**Memristive Optical Switch Matrix.**

The memristive optical switch matrix demonstrated here is a 4×4 non-blocking optical switch with Benes topology, as shown in Fig. 4a, integrating six 2×2 MZI-based PZT optical memristors cascaded in three stages and two waveguide crossings (Supplementary section **11**). Initially, these six 2×2 MZIs showed notable random phase errors due to the random variation of the waveguide dimensions, which is a common challenge in developing large-scale PICs[44]. Fig. 4b shows the measured transmissions $T_{ij}$ ($i, j$=1, 2, 3, 4) from port $I_i$ to port $O_j$ at the initial state without any calibration, illustrating that light from any one input port was routed to the target and non-target output ports with randomly power ratios (Supplementary section **12**). In this case, careful calibration is often inevitable for conventional MZIs, which introduces high complexity for the development and also consumes additional power to compensate the random phase errors[44]. Fortunately, the initial random phase errors can be compensated conveniently by utilizing the PZT non-volatility to introduce permanent phase-correction for each 2×2 MZI. By applying the appropriate setting voltage, we achieved phase shifts ranging from 0 to $\pi$, effectively reconfiguring each MZI to the desired ON or OFF state. Several typical routing configurations were demonstrated successfully by applying optimal setting voltages $V_{set}$, as summarized in Fig. 4c-e and Table S2 in the Supplemental materials. For example, we configured the routes non-volatilely as follows: (1) Configuration #1 with the routines of $I_1 \rightarrow O_1$, $I_2 \rightarrow O_2$, $I_3 \rightarrow O_3$ and $I_4 \rightarrow O_4$; (2) Configuration #2 with the routines of $I_1 \rightarrow O_3$, $I_2 \rightarrow O_4$, $I_3 \rightarrow O_1$, and $I_4 \rightarrow O_2$; (3) Configuration #3 with the routines of $I_1 \rightarrow O_2$, $I_2 \rightarrow O_1$, $I_3 \rightarrow O_3$, and $I_4 \rightarrow O_4$. All these non-volatile reconfigurations of optical routines were realized with decent extinction ratios higher than 15 dB within the wavelength range of 1530-1600 nm. In particular, for a single MZI, one can achieve non-volatile optical power routing with a low excess loss less than 0.3 dB and high extinction ratios greater than 30 dB at the center wavelength (Supplementary section **13**). In theory, the performance could be further enhanced with fine phase control. This memristive optical switch matrix was further used to demonstrate high-bit-rate data routing by choosing configuration #3 as an example (Supplementary section **13**), and 40 Gbps non-return-to-zero (NRZ) data were launched into the corresponding input port. Fig. **4f** shows the measured eye-diagrams for the data routing in the configuration #3, exhibiting high signal-to-noise ratios, showing great potential for developing large-scale N×N optical switch matrix with excellent non-volatile reconfiguration and high-speed operation. The ability to compensate the initial random phase errors non-volatilely eliminates the need for continuous power consumption for correction, representing a great advancement for developing programmable photonic chips.

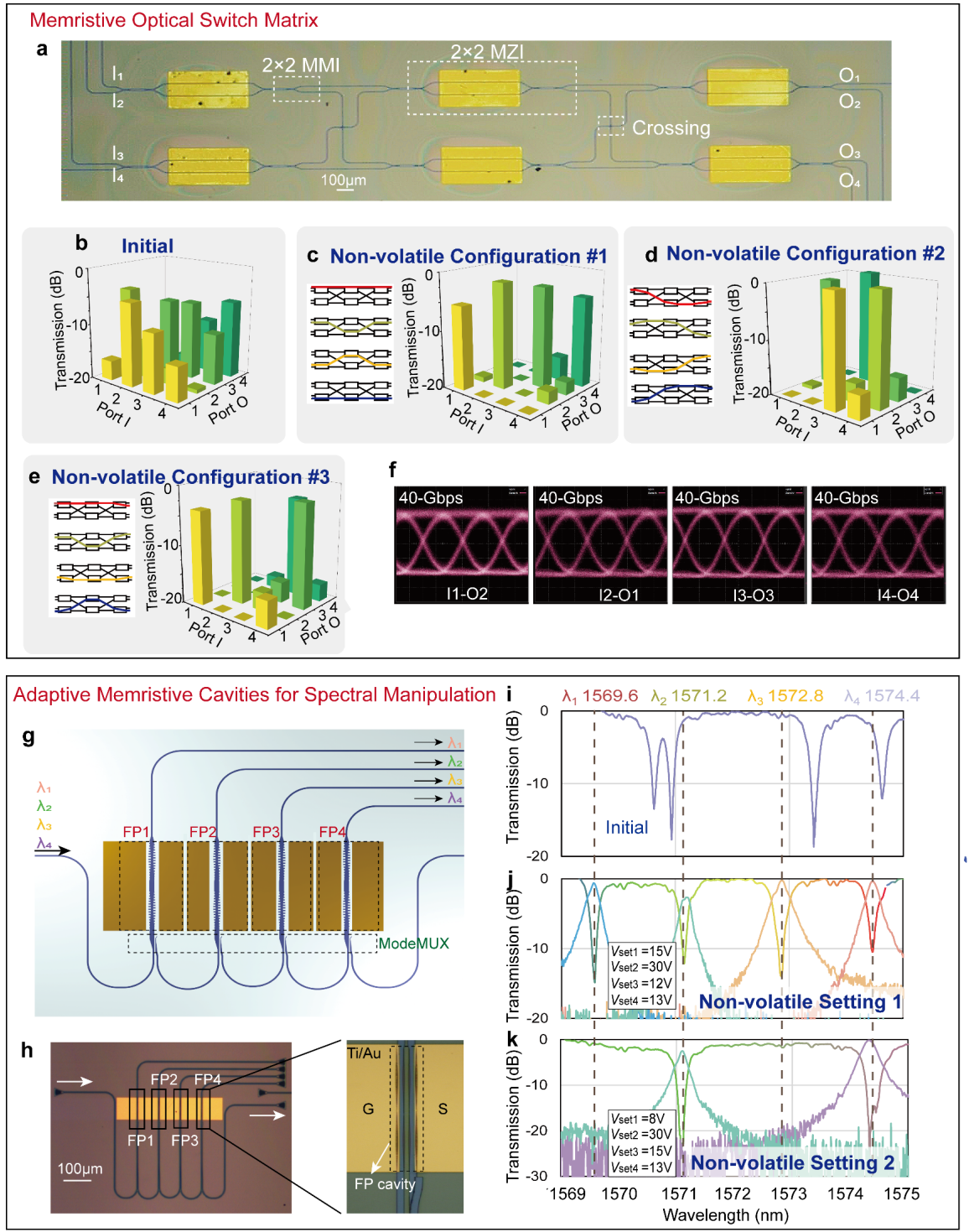

**Fig. 4 | Examples of multifunctional photonic circuits enabled by scalable PZT optical memristors: a**, Optical microscopy image of the 4×4 optical switch. **b**. Chart of the measured transmissions from the input ports ($I_1$, $I_2$, $I_3$, and $I_4$) to the output ports ($O_1$, $O_2$, $O_3$, and $O_4$) at the initial state with random phase errors. **c-e** Chart of the measured transmissions from the input ports ($I_1$, $I_2$, $I_3$, and $I_4$) to the output ports ($O_1$, $O_2$, $O_3$, and $O_4$) when all the 2×2 MZI switches are reconfigured with the desired optical routines (as defined by configurations #1, #2 and #3) by applying the appropriate setting voltages. **f**. Measured eye diagrams for the optical signals propagating along the routines ($I_1$-$O_2$, $I_2$-$O_1$, $I_3$-$O_3$, $I_4$-$O_4$) as defined by configuration #3. FP-cavity-based PZT optical memristors in cascade for adaptive memristive spectral manipulation: **g**. Schematic configuration; **h**. Optical microscopy image of the fabricated device; **i**. Measured transmission spectra at the through ports for the initial state without any correction; **j**. Measured transmission spectra at the through/cross ports when the resonance wavelength of each FP cavity was aligned carefully to have uniform channel spacing by applying the corresponding setting voltage $V_{set}$; **k**. Measured transmission spectra at the through/cross ports when channels #1 and #3 are erased and retrospectively reset so that their resonance wavelengths are respectively aligned with channels #2 and #4.

**Adaptive Memristive Optical Cavities for Spectral Manipulation.**

Optical cavities have been developed very popularly and used very widely, while the resonance wavelengths have to be tuned for alignment with additional power consumption [42]. Here, we developed an adaptive approach of non-volatilely fine-tuning the resonance wavelengths of multiple FP cavities in cascade for PZT optical memristors (Fig. 4g-h). Fig. 4i gives the measured initial transmission at the through port, showing high non-uniformity of the channel spacing, which seriously hinders the application in systems. When channels #1- #4 were tuned by applying the appropriate setting voltages $V_{set}$ of 15, 30, 12, and 13V, the resonance wavelengths were well aligned with a uniform channel spacing of 200 GHz (1.6 nm), as shown in, as shown in Fig. 4j. Note that the Q-factor changes of these FP cavities were negligible, further validating that no additional loss was introduced. To further verify the non-volatile spectral manipulation, the setting states of channels #1 and #3 were erased and reset with 8 V and 15 V, in which way their resonance wavelengths were aligned with channels #2 and #4, respectively, as shown in Fig. 4k. Accordingly, the present four-channel FP-cavity-based PZT optical memristors were reconfigured successfully to have a two-channel spectral response with a channel spacing of 400 GHz and enhanced extinction ratios of >25 dB. The chip was also characterized after it was stored at room temperature for three weeks, exhibiting a gentle resonance-wavelength shift less than 0.01 nm, thus verifying a high long-term stability (Supplementary section **14**). which is crucial for practical applications. The demonstrated non-volatility addresses the key issues for optical cavities used for the applications like wavelength-selective switches (WSSs), which are crucial components in flexible optical networks [45]. In addition, the present structure could also work as a modulator array available to work together with an optical frequency comb or multi-wavelength lasers for ultra-high capacity optical interconnects [42] or optical computing [46,47] with multiple wavelength-channels.

**Vision of potential applications of PZT optical memristors**

Our vision of potential applications of the developed PZT optical memristors is illustrated in Fig. 5, which highlights the diverse application prospects of the thin-film PZT platform for PICs, showcasing great potential in optical interconnects, microwave photonics, optical neural networks, and Lidar, etc. As shown in Fig. 5a., a general architecture of compact optical transmitters/receivers with multiple FP-cavity modulators or filters in cascade is presented, as desired for high-capacity wavelength-division-multiplexed optical interconnects. For this optical transmitter, the volatility of these cascaded FP-cavity modulators supports high-speed EO modulation, while their electrical non-volatility is used for precise trimming to align the central wavelengths as defined. For this receiver, the non-volatility is utilized for achieving accurate wavelength alignment for the multi-channel FP-cavity filter.

The present thin-film PZT platform enables various large-scale reconfigurable PICs as also. For example, as shown in Fig. 5b, the integration of multiple high-speed EO MZI switches and low-loss waveguide spirals can achieve high-speed tunable optical delaylines and the array, facilitating the realization of high-speed parallel signal processors that can manipulate signals across temporal, wavelength, and spatial dimensions. The chip can be reconfigured conveniently to support a range of advanced functionalities, including accurate microwave reception, narrowband microwave filtering, and wide-bandwidth arbitrary waveform generation for optical computing. The inherent non-volatility of PZT optical waveguides allows for precise trimming of random phase errors in MZI switches, greatly simplifying the control systems and significantly reducing the energy consumption for the PICs when used for handling complex and diverse signal processing tasks. We also present a general optical neural network (ONN) architecture in Fig. 5c, which can be implemented efficiently using the present PZT optical memristors equipped with MZIs or MRRs, without any additional complexity of design and presents significant challenges. With this ONN, high-speed training is facilitated through the volatility of the thin-film PZT MZI when the driving voltage $V_{dr}$ is below the threshold voltage $V_{th}$. After the network is trained very well, non-volatile setting can be achieved by appropriately adjusting one arm of the MZI-based optical memristors, depending on the setting voltage $V_{set}$.

The last but not the least, with the high-speed EO modulation capability of the present PZT optical memristors when working below $V_{th}$, an optical phased array (OPA) with frequency-modulated continuous-wave (FMCW) technology is achievable for all-solid-state light detection and ranging (LiDAR). When compared with the popular thermo-optically modulated silicon OPAs, the present PZT optical memristors have significant advantages of consuming ultra-low energy, eliminating the thermal crosstalk and enabling ultra-fast modulation speeds, which is especially crucial within dense OPAs for high-resolution LiDAR systems[48].

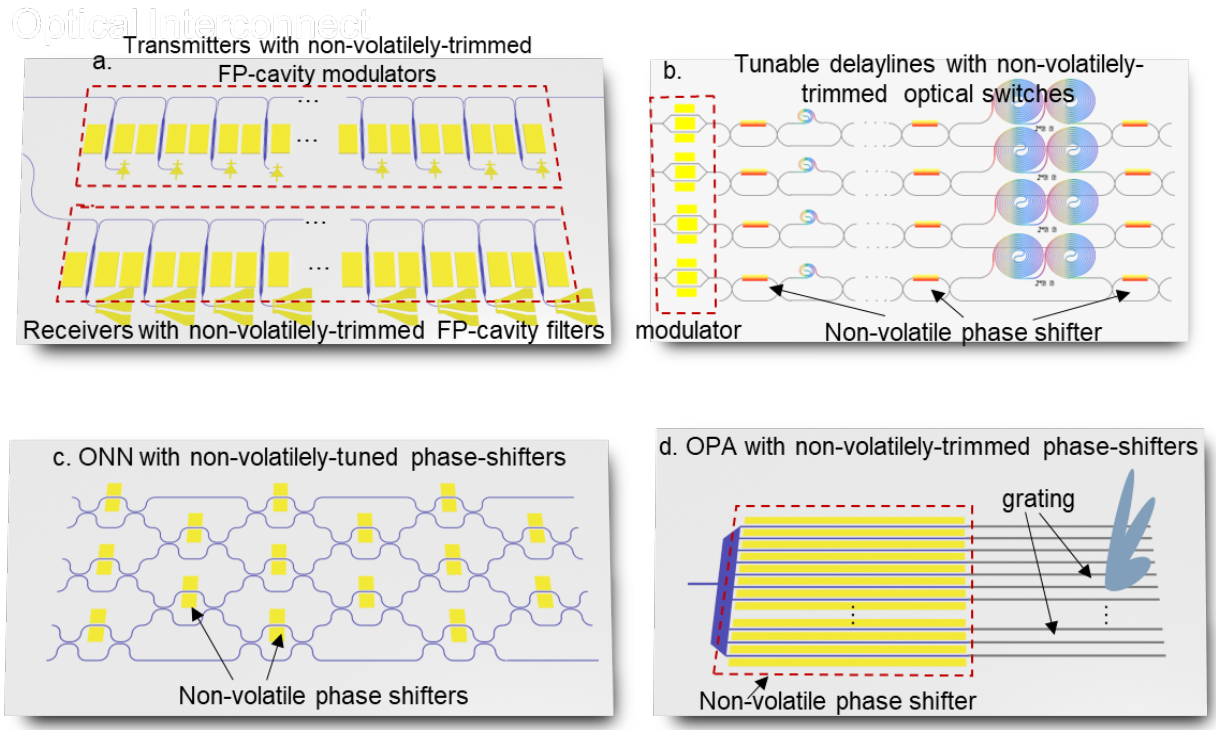

**Fig. 5 | Potential applications of our optical PZT memristors**

## Conclusions and Discussion

**Table. 1 Comparison Between electrically-controlled on-chip optical memristors.**

| Reference | [12] | [18] | [21] | [23] | [19] | This work |
|---|---|---|---|---|---|---|
| Mechanisms/materials | PCMs (e.g., $Sb_2S_3$) | MEMS | Charge | $BaTiO_3$ (BTO) | Magneto-optic | PZT |
| State Levels | 32 | 2 | 6 | >8 | 11 | **>64** |
| Non-volatile switching Energy | 56 nJ (1.7μJ [a]) | / | 0.36 pJ | 26.7 pJ | / | **0.8 pJ** |
| Switching cycles | 10 | / | 1000 | 300 | 2.4 billion | **>10000** |
| Standby power consumption | No | No | No | Yes | No | **No** |
| Propagation **Loss** dB/cm | 240 [b] | <1 | <4 [c] | 4.8 | 81[d] | **1.9** |
| Modulation depth | $\Delta n$ ~0.018, $\Delta\lambda$~0.39nm | π (phase) | $\Delta n$ ~2×10⁻³, $\Delta\lambda$~0.8nm | $\Delta n$ ~2×10⁻³, $\Delta\lambda$~0.2nm | $\Delta\lambda$~0.15nm | $\Delta n$ ~4.6×10³, $\Delta\lambda$>3.0nm |
| Retention Time | Medium | Long | Long | Long | Long | Long |
| System-level integration | No | No | No | No | No | **Yes** |
| Functional duality | No (non-volatile) | No (volatile) | No (non-volatile) | No (non-volatile: DC bias needed) | **Yes** | **Yes (non-volatile & volatile)** |
| Programming speed | ~nanosecond | / | >1 Gbps | / | 1 Gbps | **48 Gbps** |
| Key fabrication processes | Sputtering | HF-etching | Doping Bonding | Molecular beam epitaxy & Direct wafer bonding & Wafer grinding | Wafer bonding &polishing or deposition | spin-coating |

| Fabrication Complexity | High | High | High | High | High | **Low** |
| Control | Complex pulse control for heating high temperature | Moderate High voltage | Easy Electrically | Easy Electrically | Easy Electrically | Easy Electrically |

[a] Energy consumption per switching event for amorphization (crystallization).
[b] Excess loss from amorphized-PCM is usually low, while the loss of crystallized-PCM-clad waveguides is estimated to be up to ~240 dB/cm.
[c] Estimated from a 20-μm-radii MRR which has a propagation loss of <0.05 dB.
[d] Estimated from a 35-μm-radii MRR which has a propagation loss of 1.8 dB.

In Table 1, we compare the state-of-the-art demonstrations of various electrically-controlled optical memristor. PCMs are noted for their compact footprint, multi-level storage and large modulation depth but face significant challenges in cycling endurance and the complexity of pulse control needed for effective heating. MEMS structures often require large electrostatic combs, complicating their monolithic integration with other functional devices. Furthermore, the programming speed requires further enhancement to meet the demands of computing applications. BTO is capable of quasi-continuous, multi-level non-volatile control, while maintaining the state requires a continuous DC bias and leads to increased power consumption. Other optical memristors utilizing charge trapping or magneto-optic effects provide high endurance and faster programming speeds, yet there is room for improvement in modulation depth. Additionally, these optical memristors are still developed at the level of individual elements due to the fabrication complexity and propagation losses, which introduces some constraint to the scalability for multifunctional photonic integrated systems, and more efforts are demanded.

In contrast, our PZT optical memristors introduce a groundbreaking solution that supports recorded non-volatile setting levels with more than 6 bits, high cycling endurance, low non-volatile switching energy, and substantial modulation depth. A significant advancement is marked by uniquely merging non-volatile and volatile technologies in a single platform. Such capability has not been reported yet before. The functional duality opens new paradigms in high-speed and energy-efficient photonic computing/processing, where the weight array could be retrained volatilely with a fast speed of more than 48 Gbps with ultra-low energy consumption of 450 fJ/bit and high-precision non-volatilely storage. The wafer-scale sol-gel fabrication process ensures CMOS compatibility and scalability, while maintaining low propagation losses, making it ideal for large-scale on-chip photonic integration. These unparalleled capabilities of PZT optical memristors are extremely exciting with significantly reduced energy consumption and greatly enhanced processing speed, and show great potential for overcoming the von Neumann bottleneck, positioning PZT optical memristor as a transformative strategy for new paradigms in integrated photonics. Looking forward, we anticipate further integration of functionalities, like high-speed and energy-efficient optical interconnects, microwave photonics, optical phased array Lidar, quantum computing, optical neural networks, in-memory computing and brain-like architectures.

**Acknowledgements**
We acknowledge support from the National Natural Science Foundation of China (62405271, U23B2047, 62321166651, 62305294, and 92150302), Natural Science Foundation of Zhejiang Province (LD19F050001), Zhejiang Provincial Major Research and Development Program (2022C01103), Fundamental Research Funds for the Central Universities, China Postdoctoral Science Foundation (2023M733039), and Leading Innovative and Entrepreneur Team Introduction Program of Zhejiang (2021R01001),"Pioneer" and "Leading Goose" R&D Program of Zhejiang Province (2024C01112).

**Author contribution.** D.D. and C.L. conceived the idea, C.L. developed the theory of PZT optical non-volatility, C.L., T.S. and D.D. designed the PZT memristors and the chip layout, T.S., C.L., H.Y. and C.W. fabricated the chip, Y.Z and C.L. carried out the high-speed optical signal-processing of the memristors, H.C. and J.X. built up the system experiments for measuring eye-diagrams, H.Y., H.L., Z.X., Z.Y. contributed to the discussion of the fabrication and poling process, C.L. and D.D. wrote the manuscript, all authors discussed the results and contributed to the manuscript. D.D. and F.Q. supervised the project.

**Competing interests:**
The authors declare no competing interests.

**Data availability.**
Data underlying the results presented in this paper are not publicly available at this time but may be obtained from the authors upon reasonable request.

**Methods**
**Fabrication of PZT optical memristors.** For the fabrication of the present PZT optical memristors, as shown in Fig. S1, the surface of $SiO_2$/Si was chemically modified with an atomic seed layer firstly, which promotes the crystallization of PZT in the desired atomic lattice orientation. After the PZT thin-film was formed by the spin-coating technique, the substrate was heated in a rapid thermal annealing system at elevated temperatures under $O_2$ atmosphere. The spin-coating and rapid thermal annealing process were repeated several times to obtain the target film thickness of 300 nm. The details of the PZT deposition can refer to Juhe EO Tech. Co., Ltd. (www.eomaterials.com).The waveguide structure was patterned using the electron-beam lithography (EBL) or ultraviolet stepper lithography, and then was etched to a depth of 200 nm using the inductively coupled plasma process. Subsequently, a metal electrode pattern was created using the ultraviolet contact lithography, and the metal electrode consisting of 5-nm-thick Ti and 400-nm-thick Au layers was formed using the electron-beam evaporation and lift-off processes. Finally, a 1.8-μm-thick $SiO_2$ film was spin-coated on top as the upper-cladding.

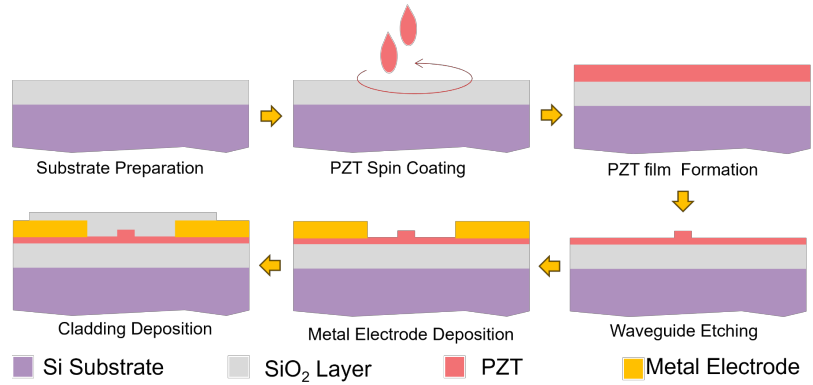

Fig. S1 Fabrication process of PZT Optical Memristors. Surface Preparation: the SiO$_2$/Si substrate was chemically modified with an atomic seed layer; PZT Deposition: PZT film was deposited by spin-coating, followed by rapid thermal annealing; a 300 nm thickness PZT film was achieved; Waveguide Patterning: the waveguide structure was patterned using electron-beam lithography (or ultraviolet stepper lithography) and etched to a 200 nm depth; Electrode Formation: metal electrodes (5-nm-thick Ti and 400-nm-thick Au) were patterned; Upper Cladding: a 1.8-μm-thick SiO$_2$ film was spin-coated.

**PZT waveguide characterization.** Fig. S2a shows the schematic diagram of the waveguide cross-section and the poling process. In this work, the thin-film PZT was poled conveniently at room temperature by applying an external electric field $E_{pol}$ of ~20 V/μm for ~30 minutes. The state of the ferroelectric domain in PZT after poling is defined as the "initial" state in this paper. The PZT optical memristors were developed on a 4-inch PZT-on-insulator wafer featuring a 300-nm-thick thin-film PZT. The PZT ridge waveguides were designed with 200-nm etching depth, while the refractive index of PZT and SiO$_2$ are given as $n_{PZT}$=2.40 and $n_{SiO2}$=1.44 at the wavelength-band of 1550 nm, respectively. The simulated field distribution for the TE$_0$ mode of the PZT optical waveguide is shown Fig. S2b, where the optical field is tightly confined in the PZT ridge with a high confinement factor over 85% due to the high refractive-index contrast, which facilitates strong electro-optic interaction in PZT.

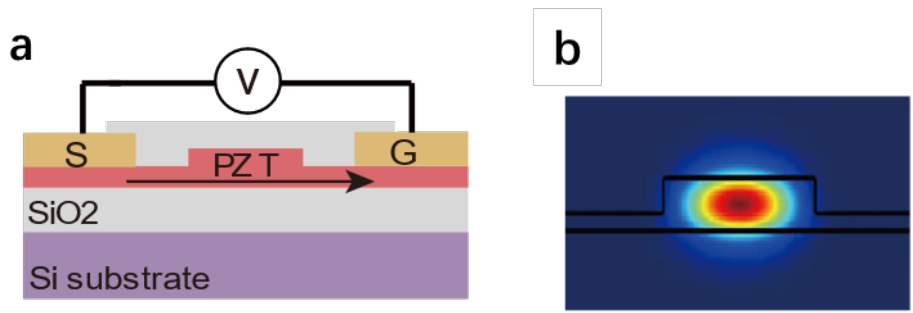

Fig. S2 PZT waveguide characterization. a. Schematic diagram of the waveguide cross-section and the poling process. b. Simulated field profile of the TE$_0$ mode.

**Experimental setup for eye diagram measurement**

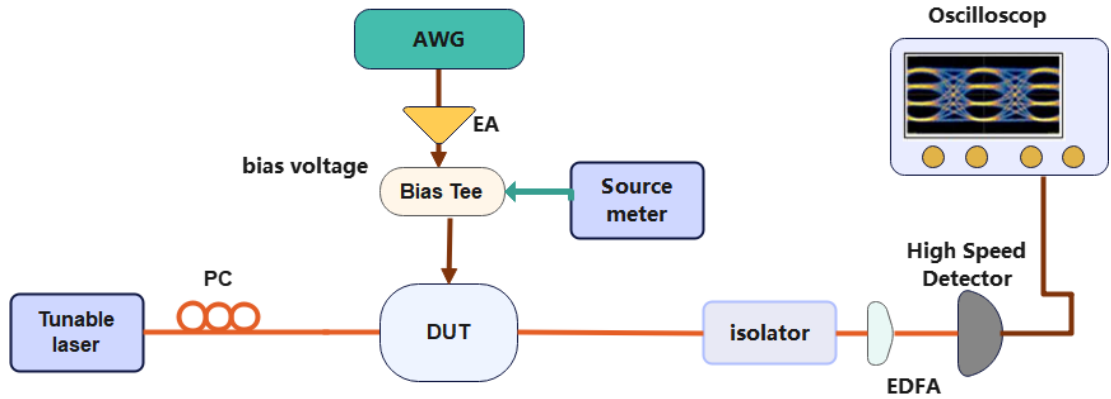

Fig. S3 Setup used for the eye diagram measurement. AWG: arbitrary waveform generator. EDFA: erbium-doped fiber amplifier. EA: electrical RF amplifier

Light was polarization-controlled and coupled to the fabricated PZT chip, and the modulator was driven by the RF signals generated by the arbitrary waveform generator (AWG) amplified by a linear electrical RF amplifier (EA), while a bias-Tee was used so that the modulator operates at the quadrature point with a given DC bias. The output optical signal was amplified using an erbium-doped fiber amplifier (EDFA) It was then detected by a 45-GHz photodetector and connected to the oscilloscope.

**Experimental setup for high-speed optical signal-processing**

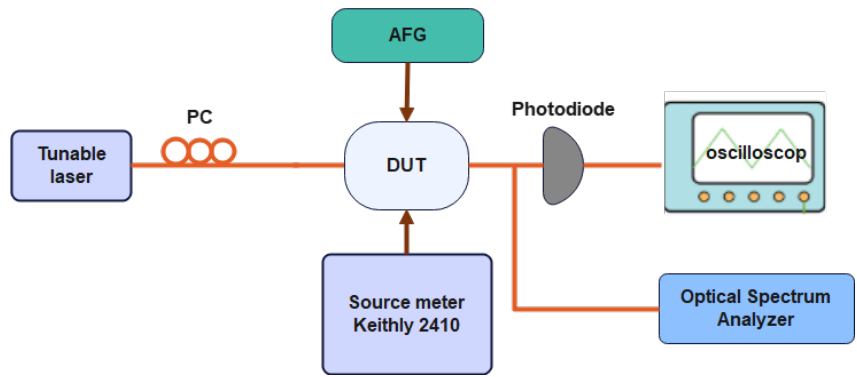

Fig. S4 Schematic of the experimental setup for high-speed optical signal-processing. AFG: arbitrary function generator, device under test (DUT).

Fig. S4 shows the high-speed optical signal-processing: a tunable laser was used as the light source and was coupled into the PZT optical memristors through a grating coupler. Here a polarization controller (PC) was used to achieve TE-polarized light for the input. A continuous rectangular-wave signal with different modulated frequency generated from an arbitrary function generator (AFG) was then applied to the DUT. The modulated optical signals were coupled from the chip to the fiber connected with a 10%: 90% power splitter and the port with 90% output-power was connected to the oscilloscope (DSO) via a photodetector (PD), while the other 10% output-power was collected by an optical spectrum analyzer, which allowed synchronized observation of the operating wavelength. A Keithley 2410 sourcemeter was employed to apply the external electric field for setting/resetting the PZT optical memristors and to measure the leakage current simultaneously.

**Experimental setup for High-bit-rate data routing**

Fig. S5 shows the measurement setup for high-bit-rate data routing with the developed memristive optical switch matrix. Here the tunable laser was set at 1550nm with an output power of 13 dBm injected into a LiNbO$_3$ Mach-Zehnder modulator. The high-speed OOK signals with a standard pseudo-random binary sequence (PRBS) pattern were generated by an arbitrary waveform generator (AWG) and then amplified by the commercial RF amplifier (SHF S807C), and finally were injected to drive the optical modulator. The modulated optical signals were then sent to the memristive optical switch matrix. At the receiving end, the optical signals were amplified by an erbium-doped fiber amplifier (EDFA,) and subsequently analyzed with a sampling oscilloscope.

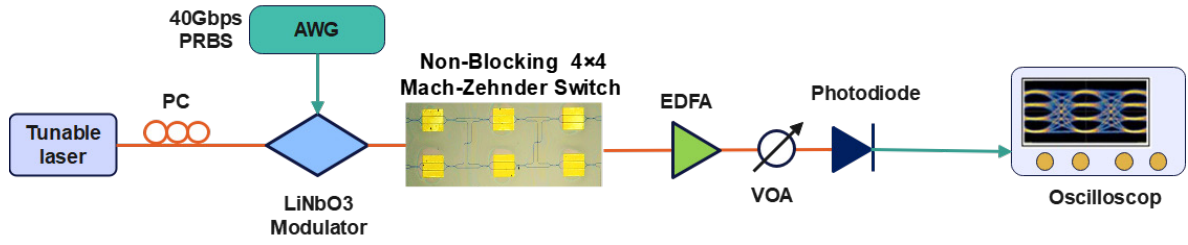

Fig. S5. Experimental setup for high-bit-rate data routing: AWG: arbitrary waveform generator. EDFA: erbium-doped fiber amplifier